\documentclass[aps,
twocolumn,
superscriptaddress,
amsfonts,amssymb,amsmath,
twoside,
letterpaper,
bibnotes,
]{revtex4}

\usepackage{bm,color,mathrsfs,graphicx}

\begin{document}
\title{
A glass-like behavior in the low-temperature specific heat \\
is a natural property of any real crystal
}

\author{A. Cano}
\email{cano@ill.fr}
\affiliation{Institut Laue-Langevin, 6 rue Jules Horowitz, B.P. 156, 38042 Grenoble, France}
\author{A. P. Levanyuk}
\affiliation{
\mbox{
Departamento de F\'\i sica de la Materia Condensada C-III, Universidad Aut\'onoma de Madrid, E-28049 Madrid, Spain
}
}
\author{S. A. Minyukov}
\affiliation{Institute of Crystallography, Russian Academy of Sciences, Leninskii Prospect 59, Moscow 117333, Russia }

\date{\today}

\begin{abstract}
We provide a rigorous calculation of the free energy of a non-metallic crystal containing a small concentration of defects. The low-temperature leading contribution is found to be $\propto T^2$. This further gives a linear-in-$T$ low-temperature specific heat as that exhibited by glasses. These results also show that, similarly to what happens in glasses, the long-wavelength spectrum of a nearly perfect crystal does not suffice to determine its low-temperature behavior.
\end{abstract}

\pacs{03.65.Yz,05.30.-d,65.40.-b,63.20.-e,63.20.Mt}

\maketitle

It is well recognized that the presence of defects may have a strong influence on the dynamics of the crystal lattice. The main effects come from the changes of the parameters characterizing the crystal and the changes in the corresponding structure induced by the defects. The former are normally modeled as local inhomogeneities, e.g., local changes in the elastic moduli and/or mass densities, while the later are encoded in the form of local fields \cite{Maradudin65,Elliott74,Teodosiu,Levanyuk_Sigov,Kosevich}. Whatever the facet of the influence of defects one considers, forced oscillations of the crystal lattice are inevitably accompanied by the (rigid) motion of defects. And this motion of defects further gives rise to the radiation of long-wavelength acoustic waves. This process is behind, for instance, the dielectric losses predicted as a result of the presence of charged defects in ionic crystals \cite{Vinogradov61}. The aim of this Letter is to show that this type of processes involving defects modeled as local fields plays a key role in the low-temperature behavior of any real crystal since any real defect has a local field component. Specifically we show that the free energy of the crystal lattice has a contribution $\propto T^2$ resulting from the presence of defects \cite{note2}, which further implies a linear-in-$T$ specific heat at low temperatures. 

In the presence of defects, there are local couplings between the modes of the perfect lattice that in general modify the corresponding density of states. These changes, however, are normally quite difficult to compute. So, to reveal the influence of defects, it is more convenient to follow a different approach. It is understood since long ago that the influence of defects can be revealed (at least partially) by considering an average over the defect configurations \cite{Elliott74,Teodosiu,Levanyuk_Sigov,Abrikosov}. Thus the influence of defects is encoded, among other quantities, into the damping of the elementary excitations of the averaged system. Then, the potential importance for the low-temperature thermodynamics of the presence of defects can be clearly foreseen by realizing that they provide a mechanism for dissipation. In fact, through the aforementioned local couplings, the energy of lattice modes of small wavelengths blur into large-wavelengths ones.

Once the elementary excitations of the system are subject to damping, their thermodynamic properties change qualitatively. For example, the free energy of a damped oscillator has a $T^2$ dependence at low temperatures \cite{Weiss}:
\begin{align}
F \approx F_0 - {\pi \over 6}{\gamma \over \omega_0^2}T^2,
\label{F_damped}\end{align}
instead of usual dependence of the undamped case \cite{Landau_SP}. Here $F_0$ is the ground-state energy, $\gamma$ is the damping coefficient and $\omega_0$ is the natural frequency of the oscillator (here and hereafter we use units such that $\hbar =k_B= 1$). In the view of Eq. \eqref{F_damped} it is clear that there will be, e.g., a linear-in-$T$ contribution to the specific heat due to these damped oscillations, as already was argued in \cite{Cano04_a,Cano04_b}. Similar speculations were made in the realm of amorphous solids \cite{Fulde71}. 
These papers deal with the damping phenomenologically and, in the case of amorphous solids, a peculiar dependence on temperature was just assumed. In the present Letter we address nearly perfect crystals and, making no assumptions about the damping, we derive these results from rigorous microscopic calculations showing that they are related to the motion of defects. 

To understand that there exists an unavoidable motion of defects that, even at zero temperature, causes the damping of phonons, let us consider an acoustic standing wave in the presence of a defect. For the sake of concreteness, we will restrict ourselves to point defects acting as dilatation centers \cite{Kosevich}. The standing wave should be imagined as a forced oscillation of the crystal whose frequency $\omega$ and wavevector $k$ are consequently independent one each other. This allows us to compute the zero-frequency limit of the damping constant for this oscillation, what is precisely the constant $\gamma$ that enters Eq. \eqref{F_damped}. As regards the defect, the gradients in pressure associated with the standing wave induce forces that make it moves \cite{Cottrell}. These forces vary in time with the frequency $\omega$ and, once the defect is in motion, this represents a source of acoustic radiation. The frequency of this radiation is again $\omega$, while its characteristic wavevector is $\omega/c$ ($c$ is the corresponding velocity of sound). Notice that, as we are considering the limit $\omega \to 0$, this later wavevector $\omega/c$ is much more smaller than the wavevector $k$ of standing wave. Already known formulas of acoustics tell us that the intensity of the acoustic radiation is $\propto \omega^2$ \cite{Landau_FM}. This energy evidently comes from the wave inducing the motion of the defect. Putting it differently, this wave losses its energy and therefore is subject to damping. And this damping is just a frequency-independent (Ohmic) damping as in Eq. \eqref{F_damped} \cite{Cano06}. It is worth mentioning that the acoustic degrees of freedom of small wavevectors then play the role of a reservoir that accepts this energy, what is in tune with the system-subsystem picture implicit in Eq. \eqref{F_damped} (see Ref. \cite{Weiss}). 

Let us proceed with more detailed calculations. To take into account the presence of dilatation centers we introduce the term
\begin{align}
- h u_{ll}
\label{F_def}
\end{align}
in the Hamiltonian density of the system, where $u_{ll}$ is the dilatation of the system. i.e., the relative volume change, and 
\begin{align}
h = \sum_a h_a \delta ({\mathbf r} - {\mathbf r}_a),
\label{}
\end{align}
with $h_a$ being a constant characterizing the change in the crystal volume caused by the $n$-th defect. As we have mentioned, the key point in our considerations is the accounting for the fact that the positions of the defects change as
\begin{align}
{\mathbf r}_a \longrightarrow {\mathbf r}_a + {\mathbf u}
\label{}
\end{align}
due to the oscillations of the lattice, where ${\mathbf u}$ is the displacement vector describing these oscillations.

We are interested in computing the temperature dependence of the free energy resulting from the thermal motion of the lattice in the presence of defects. Matsubara's method provides us a convenient diagrammatic scheme, within which the thermodynamic properties are obtained from so-called temperature Green's functions \cite{Abrikosov}:
\begin{align}
{\mathscr G}_{ij}({\mathbf r}, \tau;{\mathbf r}', \tau') = 
-{
\langle T_\tau \hat U_{i}({\mathbf r}, \tau)\hat {\overline U}_{j}({\mathbf r}', \tau'){\mathscr S} \rangle_0 
\over \langle {\mathscr S} \rangle_0} .
\end{align}
Here $T_\tau$ denotes the ordering with respect to the (imaginary time) variable $\tau \in[0,1/T]$, $\hat U_{i}$ and $\hat {\overline U}_{i}$ are Matsubara operators in the ``interaction representation'' defined from the Schr${\rm \ddot o}$dinger displacement operators $\hat u_i$ and $\hat u_i^\dag$ as
\begin{align}
\hat U_{i}({\mathbf r}, \tau) &= \sqrt{\rho }\; e^{\hat H_0\tau }\hat u_i ({\mathbf r})e^{-\hat H_0\tau },
\\
\hat {\overline U}_{i}({\mathbf r}, \tau) &= \sqrt{\rho } \; e^{\hat H_0\tau }\hat u_i^\dag ({\mathbf r})e^{-\hat H_0\tau },
\end{align}
where $ \rho $ is the density, and $\mathscr S \equiv \mathscr S (1/T)$ is the Matsubara $\mathscr S$-matrix defined as
\begin{align}
{\mathscr S}(\tau) = T_\tau \exp \left[-\int_0^\tau \hat H_\text{def}(\tau')d\tau' \right], 
\end{align}
where 
$\hat H_\text{def}(\tau) = e^{\hat H_0\tau }\hat H_\text{def }e^{-\hat H_0\tau }$. 
In deriving these formulas, the total Hamiltonian is separated into $\hat H_0 + \hat H_\text{def}$, where $\hat H_0$ is the Hamiltonian of the ideal system and $\hat H_\text{def}$ is due to the defects (we assume that the system is harmonic, so the interaction is solely due to the defects). Averaging is carried out with respect the states of the ideal system. 

The correction to free energy associated with the presence of defects can be obtained from connected diagrams contributing to the ${\mathscr S}$-matrix as
$\Delta F = -T (\langle {\mathscr S}\rangle_\text{conn} - 1)$ \cite{Abrikosov}. 
Thus, the basic ingredients in this perturbative approach are the zero-order temperature Green's functions and the defect vertices. 

In Fourier space, the temperature Green's functions for the acoustic phonons are 
\begin{align}
{\mathscr G}_{ij}^{(0)}({\mathbf k}, \omega_n) = -
{[\omega^2_n + \omega_l^2(k)] \delta_{ij} - (c_l^2-c_t^2)k_ik_j \over 
[\omega^2_n + \omega_l^2(k)] [\omega^2_n + \omega_t^2(k)] }
\end{align}
in the ideal isotropic case \cite{note_defects}. Here $\omega_n =2\pi n T$ are Matsubara frequencies ($n=0,\pm1,\pm 2,\dots$) and $\omega_{l}(k)=c_{l}k$ and $\omega_{t}(k)=c_{t}k$ are the energies of the longitudinal and transverse phonons, respectively.

To find out the factors associated with the defect vertices, let us consider the contribution to the defect Hamiltonian due to a given defect. In accordance with Eq. \eqref{F_def}, this can be written as
\begin{align}
H_{\text{def},0}= ih_0 \int{d{\mathbf k}\over (2\pi)^3} [{\mathbf k} \cdot {\mathbf u}({\mathbf k} )]e^{-i {\mathbf k} \cdot ({\mathbf r}_0 + {\mathbf u})} ,
\label{}
\end{align}
where $\mathbf {u}({\mathbf k} )$ is the Fourier transform of the acoustic displacement vector:
\begin{align}
\mathbf {u}({\mathbf r} ) = 
\int{d{\mathbf k}\over (2\pi)^3} {\mathbf u} ({\mathbf k} ) 
e^{-i {\mathbf k} \cdot {\mathbf r}}.
\label{Fourier}
\end{align}
This vector can be split into ${\mathbf u} = {\mathbf u}^\circ + {\mathbf u}' $, where ${\mathbf u}'$ represents the displacement with respect to the (static) position of equilibrium of the defect, ${\mathbf R}_0 = {\mathbf r}_0 + {\mathbf u}^\circ$, due to the oscillations of the lattice. Furthermore, by expanding $H_{\text{def},0}$ in terms of ${\mathbf u}'$, we keep the nontrivial terms of lowest order:
\begin{gather}
h_0 \int{d{\mathbf k}\over (2\pi)^3} 
[{\mathbf k} \cdot {\mathbf u}' ({\mathbf k})][{\mathbf k} \cdot {\mathbf u}'({\mathbf R}_0 )]
e^{-i {\mathbf k} \cdot {\mathbf R}_0},
\label{t1}\\
-{ih_0\over 2} 
\int{d{\mathbf k}\over (2\pi)^3} [{\mathbf k} \cdot {\mathbf u}^\circ ({\mathbf k} )]
[{\mathbf k} \cdot {\mathbf u}'({\mathbf R}_0 )]^2  
e^{-i {\mathbf k} \cdot {\mathbf R}_0 }.
\label{t2}
\end{gather}
In the second term we have the Fourier transform of the equilibrium dilatation of the crystal $-i {\mathbf k} \cdot {\mathbf u}^\circ ({\mathbf k})$. To the lowest order in $h_a$, this quantity is such that
\begin{align}
i {\mathbf k} \cdot {\mathbf u}^\circ ({\mathbf k}) 
= k_ik_jG_{ij}({\mathbf k}) \sum_a h_a e^{i {\mathbf k}\cdot {\mathbf R}_a},
\label{}
\end{align}
where $G_{ij}({\mathbf k}) = \rho^{-1}{\mathscr G}_{ij}^{(0)}({\mathbf k}, 0)$. Furthermore, taking into account Eq. \eqref{Fourier}, Eqs. \eqref{t1} and \eqref{t2} can be rewritten as
\begin{gather}
{1\over 2} \int{d{\mathbf k}\over (2\pi)^3} \int{d{\mathbf k}'\over (2\pi)^3} 
h_{ij,0}({\mathbf k},{\mathbf k}')u'_i ({\mathbf k}) u'_j({\mathbf k}'),
\label{term_1}\\
{1\over 2}\int{d{\mathbf k}\over (2\pi)^3} \int{d{\mathbf k}'\over (2\pi)^3} 
g_{ij,0}({\mathbf k},{\mathbf k}') u'_i ({\mathbf k}) u'_j({\mathbf k}').
\label{term_2}
\end{gather}
Here 
\begin{gather}
h_{ij,0}({\mathbf k},{\mathbf k}')={h_0} (k_i k_j + k_i' k_j' )e^{-i ({\mathbf k} + {\mathbf k}' )\cdot {\mathbf R}_0},
\label{}\\
g_{ij,0}({\mathbf k},{\mathbf k}') = \sum_{a}{\widetilde g_{ij,0a} }e^{-i ({\mathbf k} +{\mathbf k}')\cdot {\mathbf R}_0},
\label{}
\end{gather}
where 
\begin{align}
\widetilde g_{ij,0a} = -{h_0h_a} \int{d{\mathbf k}\over (2\pi)^3}
k_ik_jk_lk_mG_{lm}({\mathbf k}) e^{-i {\mathbf k}\cdot ({\mathbf R}_0-{\mathbf R}_a)}.
\label{}
\end{align}

\begin{figure*}[bth]
\includegraphics[width=.65\textwidth,clip]{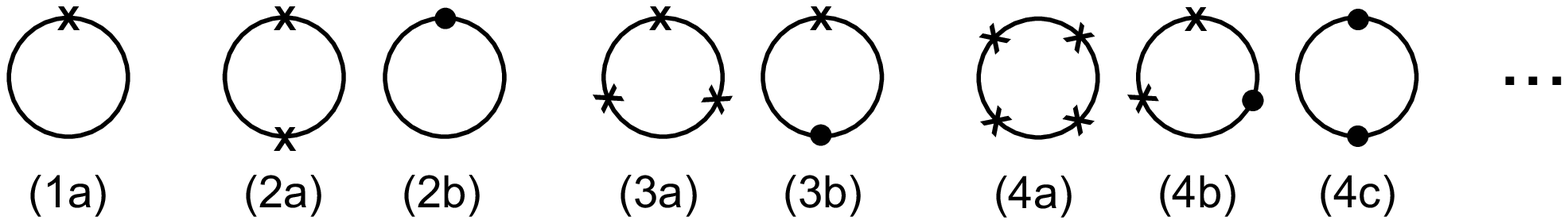}
\caption{Diagrams contributing to the free energy.}
\label{Diagrams}\end{figure*}

The terms \eqref{term_1} and \eqref{term_2} give rise to two types of vertices in the diagrams contributing to the free energy, as shown in Fig. \ref{Diagrams}. In these diagrams each line corresponds to ${\mathscr G}_{ij}^{(0)}({\mathbf k}, \omega_n) $ while defect vertices indicated by crosses are associated with the factor
\begin{align}
\rho^{-1}h_{ij,a}({\mathbf k},{\mathbf k}')
\delta_{\omega_n,\omega_{n'}},
\label{factor1}\end{align}
and defect vertices indicated by circles are associated with
\begin{align}
\rho^{-1}g_{ij,a}({\mathbf k},{\mathbf k}')
\delta_{\omega_n,\omega_{n'}},
\label{factor2}\end{align}
where $({\mathbf k},\omega_n)$ and $({\mathbf k}',\omega_{n'})$ are the arguments of the Green's functions meeting the corresponding vertex. At these vertices, sum with respect to the subindices of the Green's functions is also carried out taking into account the subindices in the above factors.

For the sake of simplicity, let us assume that defects do not change the total volume of the system and that there is no correlation between them. In this case the coefficients $h_a$ can be taken as $\pm h$, with the same concentration of defects for each sign $N_\pm = N/2$. 

The contribution of diagrams containing an odd number of cross vertices then vanish after averaging over the defect configuration. So the lowest order correction given by these diagrams is due to those of the type (2a) in Fig. \ref{Diagrams}. Furthermore, these diagrams give a nonzero contribution after averaging only if the crosses refer to the same defect. 
As a result we find a contribution proportional to
\begin{widetext}
\begin{align}
Nh'^2T\sum_{n}\int {d{\mathbf k}\over (2\pi)^3}\int{d{\mathbf k}'\over (2\pi)^3}
\left(k_ik_jk_kk_l + 2k_ik_jk_k'k_l' + k_i'k_j'k_k'k_l'\right)
{\mathscr G}^{(0)}_{il}({\mathbf k},\omega_n){\mathscr G}^{(0)}_{jk}({\mathbf k}',\omega_n),
\label{F_def_tot}\end{align}
\end{widetext}
where $h' = h/\rho$. At low temperatures, the most important term comes from the first product of $k$'s in the integrand. This term reduces to
\begin{align}
-{Nh'^2 T}\sum_{n}\int {d{\mathbf k}\over (2\pi)^3}\int{d{\mathbf k}'\over (2\pi)^3}
{3k^2k^2_x\over \omega_n^2 + \omega_l^2(k)}{\mathscr G}^{(0)}_{xx}({\mathbf k}',\omega_n) .
\label{}\end{align}
Here we can write 
\begin{align}
&\sum_{n} {T\over \omega_n^2 + \omega_l^2(k)} {\mathscr G}^{(0)}_{xx}({\mathbf k}',\omega_n) 
= 
{\omega_l(k) - \omega_l(k')\over 2\omega_l(k)\omega_l(k')[\omega_l^2(k) - \omega_l^2(k')]}
\nonumber \\ &\qquad 
+
{1\over \omega_l^2(k') - \omega_l^2(k)}
\left(
{n[\omega_l(k)]\over \omega_l(k)}-{n[\omega_l(k')]\over \omega_l(k')}\right) 
\nonumber \\ &\qquad + T\sum_{n}{(c_l^2-c_t^2)(k'^2-k_x'^2)\over [\omega_n^2 + \omega_l^2(k)][\omega_n^2 + \omega_l^2(k')][\omega_n^2 + \omega_t^2(k')]},
\label{sum_n}\end{align}
where $n(\omega)=[\exp(\omega/T)-1]^{-1}$ is the Bose-Einstein distribution function. The first term at the rhs of \eqref{sum_n} does not depend on the temperature. At low temperatures, the contribution of the second term can be estimated as
\begin{align}
&\simeq 
-2Nh'^2 \int {d{\mathbf k}\over (2\pi)^3}\int{d{\mathbf k}'\over (2\pi)^3}
{k^2k_x^2 n[\omega_l(k')]\over \omega_l^2(k)\omega_l(k')}
\nonumber \\&\approx 
-2Nh'^2 {k_\text{max}^5\over c^5_l}T^2,
\label{F_result}\end{align}
where $k_\text{max}$ is the radius of the Brillouin zone. One can see that both the remaining sum in Eq. \eqref{sum_n} and the term \eqref{term_2} give similar contributions $\sim T^2$ to the free energy. In the later case, a nonzero contribution is already obtained from diagrams of type (2b) in Fig. \ref{Diagrams} [notice that the factor \eqref{factor2} already is $\sim h^2$].

At low enough temperatures, the $T^2$ contribution we have obtained due to the presence of defects can surpass the one of the ideal lattice ($\propto T^4$, see e.g. Ref. \cite{Landau_SP}). Eq. \eqref{F_result} indeed coincides with the free energy estimated for a set of harmonic oscillators, with the dispersion law $\omega_l(k) = c_lk$ and a damping constant $\gamma \sim Nh'^2 k_\text{max}^4/c_l^3$ [see Eq. \eqref{F_damped}]. As we have argued, this coincidence is not accidental. The phonon lifetime indeed can be computed within a perturbative approach similar to employed here. As a result one finds that, for large enough wavevectors, acoustic phonons are precisely subject to an Ohmic damping due to the motion of defects \cite{Cano06}. The reasonings given in Refs. \cite{Cano04_a,Cano04_b} are therefore applicable when estimating the crossover temperature. This temperature is found to be attainable experimentally (up to $\sim 1\;$K for moderate concentrations of defects).

It is worth mentioning that, as a result of higher order terms in the above expansion of the ${\mathscr S}$-matrix, one simply obtains corrections to the coefficient of the $T^2$ term in the free energy (also new terms with higher powers of $T$, unimportant at low enough temperatures). When it comes to the comparison between theory and experiments, this implies that the lowest-order result we have found is robust (qualitative agreement is expected). But it also have a more profound reading. The fact that the density of defects can be increased without altering the asymptotic low-temperature behavior allows one to speculate that the result is valid even in a glass limit (i.e., for large concentrations of defects). Indeed, already with a small concentration of defects we obtain, e.g., a linear-in-$T$ low-temperature specific heat as that exhibited by glasses (see e.g. Ref. \cite{Phillips87}). This linear-in-$T$ specific heat is normally considered as a fingerprint of glasses. However, we see that these distinctive features can already be found within the conventional scenario of a crystal with a not necessarily large concentration of defects. 

We have so far restricted ourselves to reveal the influence of defects through the local dilatation they induce. Real defects, however, have a broader influence \cite{Cano06}. Any impurity rarely acts as a dilatation center exclusively. On the contrary, they generally induce complicated force distributions. Besides defects may be individually charged and/or they can act as local electric multipoles and, consequenly, they will move due to electric fields varying in time (even inhomogeneous ones). Beyond that, the properties of the corresponding crystal can be affected by defects as more general multipoles, say ``optic multipoles'' \cite{Levanyuk_Sigov,Cano06}, that also will oscillate if there exists a wave of the corresponding optic coordinate. 
So the elementary excitations of the system inevitably will induce the motion of the defects. And this motion will generate a long-wavelength acoustic radiation. The process on which our results base is therefore universal in the sense that takes place for any type of defect. Consequently, the same thermodynamics is expected for any real crystal.

Let us mention, in closing, that the contribution we have found associated with the motion of defects is not a mere correction of previous results. It indeed may be the leading contribution by virtue of the low temperature of the system. Moreover, contrary to what happens in ideal case, large wavevectors are involved in this leading contribution \cite{note4}. This represents a deep change with respect to the paradigm for the low-temperature behaviors (see e.g. Ref. \cite{Landau_SP,Abrikosov}), according to which these behaviors are associated with the macroscopic properties of the corresponding system. In some sense, it takes place the same situation as in glasses. That is, universal low-temperature behaviors are found but cannot be understood as uniquely due to the macroscopic properties of the system.

A.C. was supported by a posdoctoral fellowship from Fundaci\' on Ram\' on Areces. A.P.L. was supported by Spanish MEC (MAT2006-07196). S.A.M. was supported by the Russian Foundation for Basic Researches (RFBR), Grant 05-02-17565.



\begin{references}


\bibitem{Maradudin65} A.A. Maradudin, Rep. Prog. Phys. {\bf 28}, 331 (1965); I.M. Lifshitz and A.M. Kosevich, {\it ibid.} {\bf 29}, 217 (1966).

\bibitem{Elliott74} R.J. Elliott, J.A. Krumhansl and P.L. Leath, Rev. Mod. Phys. {\bf 3}, 465 (1974). 

\bibitem{Teodosiu} C. Teodosiu, {\it Elastic Models of Crystal Defcts} (Springer-Verlag, NY, 1982).

\bibitem{Levanyuk_Sigov} A.P. Levanyuk and A.S. Sigov, {\it Defects and Structural Phase Transitions} (Gordon{\&}Breach, NY, 1987); Phase Transit. {\bf 11}, 91 (1988).

\bibitem{Kosevich} A.M. Kosevich, {\it The Crystal Lattice} (Wiley-VCH, Weinheim, 2005).

\bibitem{Vinogradov61} V.S. Vinogradov, Fiz. Tverd. Tela {\bf 2}, 2622 (1960) [Sov. Phys.-- Solid State {\bf 2}, 2338 (1961)]; E. Schl$\rm \ddot o$mann, Phys. Rev. {\bf 135}, A413 (1964); B.M. Garin, Fiz. Tverd. Tela {\bf 32}, 3314 (1990).

\bibitem{note2} Let us stress that we are dealing with effects due to the \emph{presence} of defects, but not due to the defects themselves (i.e., not due to their internal degrees of freedom).

\bibitem{Abrikosov} A.A. Abrikosov, L.P. Gorkov and I.E. Dzyaloshinski, {\it Methods of Quantum Field Theory in Satistical Physics} (Dover, NY, 1975).

\bibitem{Weiss} U. Weiss, {\it Quantum Dissipative Systems} (World Scientific, Singapore, 1999). 

\bibitem{Landau_SP} L.D. Landau and E.M. Lifshitz, {\it Statistial Physics} (Pergamon, Oxford, 1980).

\bibitem{Cano04_a} A. Cano and A.P. Levanyuk, Phys. Rev. B {\bf 70}, 212301 (2004), cond-mat/0404063.

\bibitem{Cano04_b} A. Cano and A.P. Levanyuk, Phys. Rev. Lett. {\bf 93}, 245902 (2004), cond-mat/0404437; {\it ibid.} {\bf 96}, 039604 (2006).

\bibitem{Fulde71} P. Fulde and M. Wagner, Phys. Rev. Lett. {\bf 27}, 1280 (1971); M. Turlakov, {\it ibd.} {\bf 93}, 035501 (2004). 

\bibitem{Cottrell} This type of forces are responsible for the Cottrell mechanism that determines the concentration of solute atoms in the presence of dislocations [A.H. Cottrell and B.A. Bilby, Proc. Phys. Soc. Lond. B{\bf 62}, 229 (1949); J.P. Hirth and J. Lothe, {\it Theory of dislocations} (McGraw-Hill, 1982).] 

\bibitem{Landau_FM} L.D. Landau and E.M. Lifshitz, {\it Fluid Mechanics} (Pergamon, Oxford, 1987).
The relevant problem to our case is that of the translatory oscillations of a sphere with a fixed acceleration due to an external harmonic force (instead of oscillations with fixed amplitude or velocity).

\bibitem{Cano06} A. Cano, A.P. Levanyuk and S.A. Minyukov, cond-mat/0603343.

\bibitem{note_defects} Strictly speaking, we are modeling the defects in such a way that they will not displace if the medium is isotropic. The reason is that, in this case, dilatation centers induce pure shear deformations that vanish in average. So, in reality, they are not sensitive to, e.g., gradients of pressure if the medium is isotropic. Nevertheless, their rigid motion in anisotropic media can be reproduced even by simplifying the response of the medium and assuming that it is isotropic.

\bibitem{Phillips87} W.A. Phillips, Rep. Prog. Phys. {\bf 50}, 1657 (1987).

\bibitem{note4} In Eq. \eqref{F_result}, temperature acts as a cutoff for the integral over ${\mathbf k}'$, but not for the remaining integral over ${\mathbf k}$.

\end{references}
\end{document}